\documentclass{aastex701}

\usepackage{comment}

\newcommand{\lya}{Ly$\alpha$}

\begin{document}

\title{Validating $z\gtrsim 7.5$ Lyman Break Galaxy candidates in the COSMOS field with JWST/PASSAGE}

\author[0009-0003-4658-8874]{Zoke W. Sackih}
\affiliation{Minnesota Institute for Astrophysics, University of Minnesota, 116 Church Street SE, Minneapolis, MN 55455, USA}
\email[show]{sacki014@umn.edu}  

\author[orcid=0009-0002-9932-4461]{Mason S. Huberty}
\affiliation{Minnesota Institute for Astrophysics, University of Minnesota, 116 Church Street SE, Minneapolis, MN 55455, USA}
\email{huber458@umn.edu}  

\author[0000-0002-9136-8876]{Claudia Scarlata}
\affiliation{Minnesota Institute for Astrophysics, University of Minnesota, 116 Church Street SE, Minneapolis, MN 55455, USA}
\email{mscarlat@umn.edu} 

\author[0000-0003-3108-0624]{Peter J. Watson}
\affiliation{INAF, Osservatorio Astronomico di Padova, Vicolo dell'Osservatorio 5, 35122 Padova, Italy}
\email{peter.watson@inaf.it}

\author[0000-0003-4569-2285]{Andrew J. Battisti}
\affiliation{International Centre for Radio Astronomy Research, University of Western Australia, 7 Fairway, Crawley, WA 6009, Australia}
\affiliation{Research School of Astronomy and Astrophysics, Australian National University, Cotter Road, Weston Creek, ACT 2611, Australia}
\email{andrew.battisti@anu.edu.au}

\author[0000-0002-0072-0281]{Farhanul Hasan}
\affiliation{Space Telescope Science Institute, 3700 San Martin Drive, Baltimore, MD 21218, USA}
\email{fhasan@stsci.edu}

\author[0000-0001-8587-218X]{Matthew J. Hayes}
\affiliation{Stockholm University, Department of Astronomy, AlbaNova University Center, SE-106 91 Stockholm, Sweden}
\email{matthew@astro.su.se} 

\author[0000-0001-6919-1237]{Matthew A. Malkan}
\affiliation{University of California, Los Angeles, Department of Physics and Astronomy, 430 Portola Plaza, Los Angeles, CA 90095, USA}
\email{malkan@astro.ucla.edu}

\author[0000-0001-7166-6035]{Vihang Mehta}
\affiliation{IPAC, California Institute of Technology, 1200 E.~California Blvd, Pasadena, CA 91125, USA}
\email{vmehta@ipac.caltech.edu}

\author[0000-0001-5294-8002]{Kalina~V.~Nedkova}
\affiliation{IPAC, California Institute of Technology, 1200 E.~California Blvd, Pasadena, CA 91125, USA}
\affiliation{Department of Physics and Astronomy, Johns Hopkins University, 3400 North Charles Street, Baltimore, MD 21218, USA}
\affiliation{Space Telescope Science Institute, 3700 San Martin Drive, Baltimore, MD 21218, USA}
\email{knedkova@stsci.edu}

\author[0000-0002-9946-4731]{Marc Rafelski}
\affiliation{Space Telescope Science Institute, 3700 San Martin Drive, Baltimore, MD 21218, USA}
\email{mrafelski@stsci.edu}

\author[0000-0003-0980-1499]{Benedetta Vulcani}
\affiliation{INAF, Osservatorio Astronomico di Padova, Vicolo dell'Osservatorio 5, 35122 Padova, Italy}
\email{benedetta.vulcani@inaf.it}


\begin{abstract}
    We analyze spectroscopy from one NIRISS pointing in the JWST-PASSAGE program for seven candidate $z \gtrsim 7.5$ photometrically-selected COSMOS-Web sources.
    We spectroscopically confirm one out of seven sources as a Lyman break galaxy (LBG) at $z=7.962^{+0.003}_{-0.006}$, with $m_{F150W} = 25.9$ (AB). The remaining sources are too faint in the continuum (i.e., $m_{F150W} \gtrsim 26$ AB) to provide a redshift measurement from the Lyman break, and do not show emission lines in their spectra. Although this study contains only one spectroscopically confirmed source, the confirmation of a luminous $z \sim 8$ galaxy within this $\sim4.8$ arcmin$^2$ field implies a surface density of $\sim 0.21^{+0.59}_{-0.17}$ arcmin$^{-2}$, $\approx 10\times$ higher than inferred from wide-area photometric surveys, suggesting a potential overdensity at $z\sim8$ in the COSMOS field.
\end{abstract}

\section{Introduction}

A critical period in the early universe was the Epoch of Reionization (EoR), during which the first galaxies and active galactic nuclei (AGN) ionized the diffuse intergalactic medium. However, studies of the EoR remain limited by the small number of known objects at $z\gtrsim6$, where it is observed to end \citep[e.g.,][]{umeda2024}. While JWST has enabled spectroscopic redshift measurements of galaxies out to $z\approx 14$ \citep[e.g.,][]{naidu2026}, its deep pencil-beam surveys probe small sky areas and are insensitive to rare, bright sources. This problem is exacerbated by their strong clustering, which introduces significant field-to-field variance. To overcome this, the Parallel Application of Slitless Spectroscopic to Analyze Galaxy Evolution survey \citep[PASSAGE,][]{malkan2025} observed 63 fields across the sky, mitigating cosmic variance and improving the detection of rare, bright sources. These sources enable precise redshift measurements via the Lyman break
and provide high signal-to-noise spectra for detailed study.

In this research note, we analyze a PASSAGE field within the COSMOS footprint, an excellent region for identifying bright $z\gtrsim 6$ galaxies due to its large area and extensive multi-wavelength coverage \citep{scoville2007,weaver2022,shuntov2025}. We use COSMOS-Web, the most recent compilation of photometric redshifts in COSMOS \citep{shuntov2025}, to select candidate galaxies, and leverage PASSAGE data to validate the redshifts of sources at $z\gtrsim 7.5$.
We use $\Omega_{\rm m}=0.31$, $\Omega_{\rm \Lambda}=0.69$, $H_0=67.7~{\rm km/s/Mpc}$ \citet[][]{planck2020} and the AB magnitude system \citep{oke1983}.



\section{Sample Selection and Spectroscopic Observations}


From the COSMOS-Web catalog we identify 2048 galaxies for which the photometric redshift\footnote{We use $z_{\mathrm{phot}}$, the photometric redshift that minimizes $\chi^2$. $z_{16}$ and $z_{84}$ are the 16th and 84th redshift percentiles respectively.} is $z_{\mathrm{phot}} > 7.5$, $m_{F150W} < 28$, and $z_{84} - z_{16} < 1$  \citep[ensuring a single narrow peak in the redshift probability distribution, ][]{arnouts2002,ilbert2006}.
We cross-match this sample with the PASSAGE field Par028 \citep{huberty2026}, the deepest field in PASSAGE and observed in three NIRISS grism filters (F115W, F150W, and F200W), covering $\sim$4.8~arcmin$^2$. This field is well suited for identifying  galaxies at $z \gtrsim 7.5$, where the Lyman break enters the F115W bandpass.


\section{Analysis and Results}
\begin{figure*}[h]
    \centering
    \includegraphics[width=.97\linewidth]{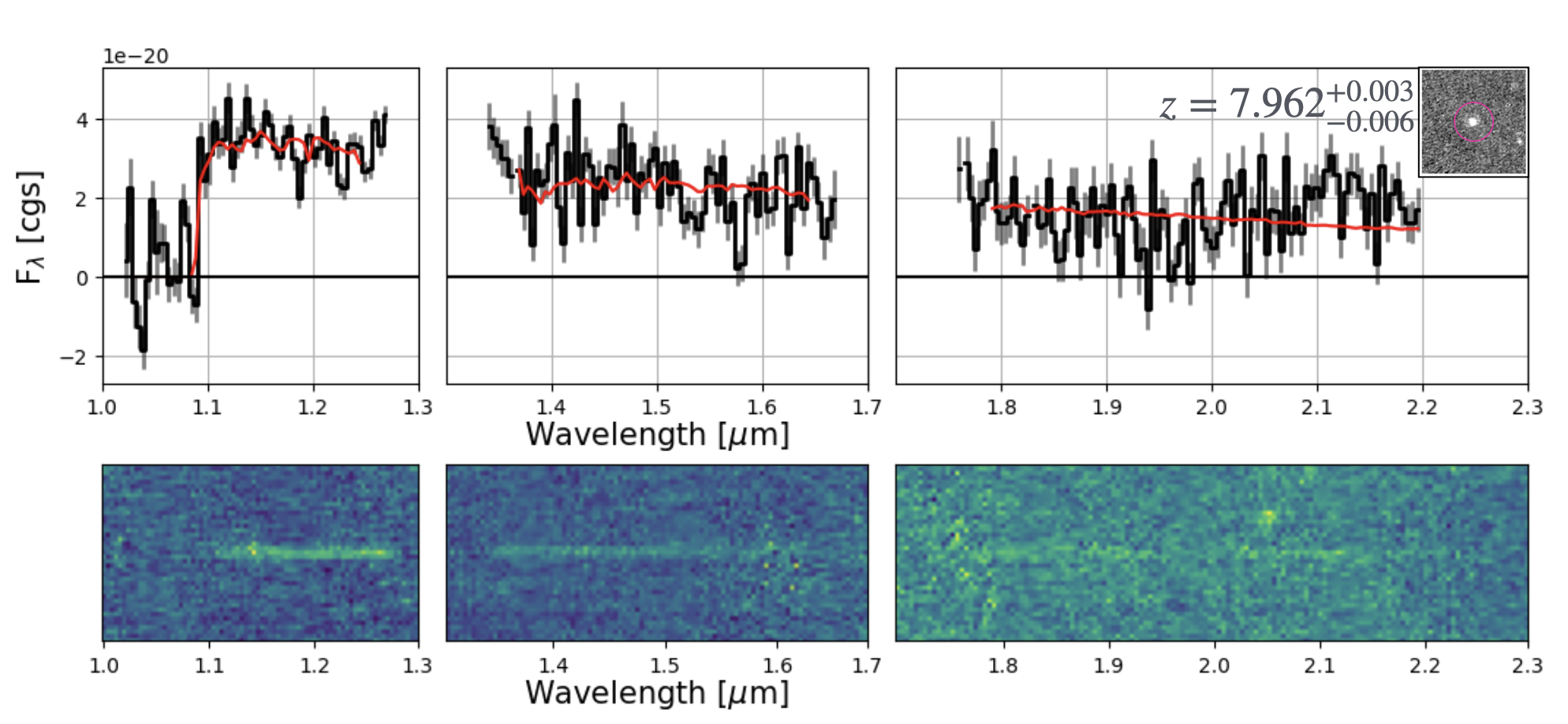}
    \caption{Spectroscopic confirmation and spectral modeling of the sole confirmed LBG in Par028.
\textit{Top panels:} Observed NIR spectrum (black) with $1\sigma$ uncertainties (gray), overlaid with the best-fit model (red), highlighting a pronounced spectral break at $z = 7.962^{+0.003}_{-0.006}$. The inset shows a $4'' \times 4''$ cutout of this source.
\textit{Bottom panels:} Two-dimensional spectra over the corresponding wavelength ranges, showing continuum emission and no evidence for low-redshift emission-line interlopers. 
\label{fig:llbg}}
\end{figure*}

We identify seven sources matched within $0\farcs3$ between PASSAGE Par028 and COSMOS-Web. Visual inspection of the 1D and 2D spectra reveals a clear Lyman break in one galaxy with $z_{\mathrm{phot}} = 8.33$ (Figure~\ref{fig:llbg}). This source, the brightest in Par028 ($m_{F150W} = 25.9$~AB), is shown with its 1D (top) and 2D (bottom) spectra. The remaining six sources show no detectable continuum or spectral features, precluding spectroscopic redshift measurements.
We model the confirmed LBG spectrum using the Bayesian Analysis of Galaxies for Physical Inference and Parameter Estimation \citep[BAGPIPES;][]{carnall2018,carnall2019}, adopting an exponential star formation history. 
BAGPIPES yields a best-fit redshift of $z = 7.962^{+0.003}_{-0.006}$; however, a potential damped \lya\ absorber may bias this estimate slightly high.

We compare our result to the surface density of $z \sim 8$ candidates from the SuperBorg survey \citep{leethochawalit2023,SuperBoRG}, which covers 1053~arcmin$^2$ and identified 26 photometric $z \sim 8$ candidates down to $H_{160} < 26.6$, corresponding to $\sim 0.025$ candidates per arcmin$^2$. In contrast, within the $2.2' \times 2.2'$ Par028 field, we identify one spectroscopically confirmed source, or $\sim 0.21^{+0.59}_{-0.17}$ arcmin$^{-2}$ \citep[with a Poisson uncertainty following][]{Gehrels1986}. While the uncertainty is large due to small-number statistics, this comparison suggests the COSMOS field may host an elevated surface density of $z \approx 8$ galaxies relative to wide-area photometric surveys, potentially reflecting cosmic variance.

\section{Conclusions}
We use JWST/PASSAGE slitless spectroscopy to validate COSMOS-Web photometric $z \gtrsim 7.5$ candidates. Of seven candidates, we securely confirm one bright LBG at $z = 7.962^{+0.003}_{-0.006}$; the remainder are too faint for continuum-based confirmation and show no emission lines. The inferred surface density, though highly uncertain due to small-number statistics, tentatively suggests an overdensity of luminous $z \sim 8$ galaxies in this field, potentially reflecting cosmic variance.

\section{Acknowledgement}
This research was supported by the International Space Science Institute in Bern, through  International Team project \#24-624. We acknowledge support by NASA through grant No. JWST-GO-1571, and by the University of Minnesota Undergraduate Research Opportunities Program Grant. All the JWST data used in this paper can be found in MAST: \href{https://archive.stsci.edu/doi/resolve/resolve.html?doi=10.17909/h60a-9809}{doi:10.17909/h60a-9809}.

\bibliography{refrences}{}
\bibliographystyle{aasjournalv7}


\end{document}